\documentstyle[aps,amsfonts]{revtex}

\def\ZR{{{\mathbb Z}}}

\def\K{{\cal K}}

\def\M{{\cal M}}
\def\N{{\cal N}}

\def\R{{\cal R}}

\def\SHB{{\mathbb S}}
\def\para{{\scriptscriptstyle /\!/}}
\def\Tdot#1{{{#1}^{\hbox{.}}}}
\def\Tddot#1{{{#1}^{\hbox{..}}}}
\def\Frac(#1/#2){\left(\frac{#1}{#2}\right)}
\def\Order#1{{\rm O}\!\left(#1\right)}

\begin{document}
\title{Behavior of Cosmological Perturbations \\
in the Brane-World Model%
\footnote{A talk in the conference CAPP2000 at Verbier, Switzerland,
July 18 to 27, 2000, to be published in the proceedings ``Cosmology and 
Particle Physics'', ed. J. Garcia-Bellido, R. Durrer and
M. Shaposhnikov (AIP)}}

\author{Hideo Kodama}
\address{
Yukawa Institute for Theoretical Physics\\
Kyoto University, Kyoto 606-8502, Japan
}

\maketitle

\begin{abstract}
In this paper we present a gauge-invariant formalism for 
perturbations of the brane-world model developed by the author, A. 
Ishibashi and O. Seto recently, and analyze the behavior of 
cosmological perturbations in a spatially flat expanding universe 
realized as a boundary 3-brane in AdS$^5$ in terms of this 
formalism. For simplicity we restrict arguments to scalar 
perturbations. We show that the behavior of cosmological 
perturbations on superhorizon scales in the brane-world model is the same 
as that in the standard no-extradimension model, irrespective of the 
initial condition for bulk perturbations, in the late stage when the 
cosmic expansion rate $H$ is smaller than the inverse of the bulk 
curvature scale $\ell$. Further, we give rough estimates which 
indicate that in the early universe when $H$ is much larger than 
$1/\ell$, perturbations in these two models behave quite differently, and 
the conservation of the Bardeen parameter does not hold for 
superhorizon perturbations in the brane-world model.
\end{abstract}

\section{Introduction}

Recently, motivated by the $S^1/\ZR_2$ reduction of the M-theory and 
the hierarchy problem in particle physics, a fascinating model was 
proposed by Randall and Sundrum, in which 4-dimensional Minkowski 
spacetime is realized as a boundary 3-brane in the 5-dimensional 
anti-de Sitter spacetime%
\cite{Randall.L&Sundrum}.
The most surprising feature of this model is that the 5-dimensional 
gravity is confined in the brane in the sense that the standard 
Newtonian law of gravity holds in the non-relativistic limit on 
scales much larger than the bulk curvature scale $\ell$, and the 
massive Kaluza-Klein modes of gravitational waves decouple from the 
massless mode and their mode functions are suppressed near the brane.

After the paper by Randall and Sundrum, brane-world cosmology has 
been actively studied by many people, and it has been shown that any 
spatially homogeneous and isotropic universe can be embedded as 
3-brane in AdS$^5$. Further, in order to see whether the brane-world 
model can provide a new framework for cosmology consistent with 
observations, formalisms to investigate behavior of perturbations in 
the brane-world model have been proposed by various people%
\cite{%
Mukohyama.S,%
Maartens.R2000A,%
vandeBruck.C&&2000A,%
Koyama.K&Soda2000A,%
Langlois.D&Maartens&Wands2000A}.
In particular, the author developed with A. Ishibashi and O. Seto a 
complete gauge-invariant formalism for perturbations in the 
brane-world model in which the Einstein equations for perturbations 
are reduced to a single master equation with a boundary condition on 
a dynamical brane%
\cite{Kodama.H&Ishibashi&Seto2000A}. In the present paper, we first 
briefly describe the main point of this formalism specialized to 
scalar perturbations, and then address the important issue whether 
the behavior of scalar perturbations in the brane-world model is the 
same as that in the conventional no-extradimension model. 

\section{Gauge-Invariant Formalism of Perturbations}

In this section we give the gauge-invariant formalism for scalar 
perturbations in the $(n+2)$-dimensional brane-world model in 
which the bulk spacetime $(\M,\bar g_{MN})$ is empty and its 
unperturbed background is given by $AdS^{n+2}$ with the negative 
cosmological constant $\Lambda=-n(n+1)/(2\ell^2)$. The starting 
point is the Einstein equations for the bulk,
$
\bar G_{MN}+{\bar\Lambda} \bar g_{MN}=0
$,
and Israel's junction condition for $\ZR_2$ symmetry at the 
$n$-brane $\Sigma$,
\begin{equation}
{\bar\kappa}^2 T^\mu_\nu=2(K^\mu_\nu -K\delta^\mu_\nu),
\label{JunctionCondition}
\end{equation}
where $\bar\kappa^2$ is the gravitational constant for the bulk, 
$T^\mu_\nu$ is the energy-momentum tensor of the brane, and 
$K^\mu_\nu$ is the extrinsic curvature of the brane.

\subsection{Unperturbed Background}

We assume that the unperturbed configuration including the brane is 
invariant under the isometry group $G_n$ isomorphic to the symmetry group 
of the $n$-dimensional constant curvature space $\K^n$, and that the the 
bulk manifold $\M^{n+2}$ is written as 
$\N^2\times\K^n$, where $\N^2$ is the 2-dimensional orbit space:
$
\M^{n+2}=\N^2\times \K^n \ni (y^a, x^i)=(z^M)
$.
Under this decomposition the bulk background metric takes the form
\begin{equation}
d{\bar s}^2=\bar g_{MN}dz^Mdz^N=g_{ab}(y)dy^a dy^b + r^2(y)d\sigma_n^2;
\end{equation}
where $d\sigma_n^2$ is the metric of $\K^n$ with a constant 
curvature $K$. 

The unperturbed structure of the brane is described  by the 
Robertson-Walker metric
$
ds^2=-d\tau^2+a(\tau)^2d\sigma_n^2
$,
and the energy-momentum tensor 
$
T_{\tau\tau}=\rho, 
T_{\tau i}=0,
T^i_j=p \delta^i_j
$.
The embedding of this brane into the bulk background 
spacetime is represented by functions $y^a(\tau)$ such that 
$a(\tau)=r(y(\tau))$ and $d\tau^2=-g_{ab}dy^a dy^b$. The 
junction condition is expressed as
\begin{equation}
\frac{D_\perp 
r}{r}=-\frac{{\bar\kappa}^2}{2n}\rho,\quad
(n-1)\frac{D_\perp 
r}{r}-K^\tau_\tau=\frac{{\bar\kappa}^2}{2}p,
\label{JC:BG}
\end{equation}
where $D$ is the covariant derivative with respect to the metric 
$g_{ab}(y)$ of $\N^2$ and $D_\perp=n^aD_a$ with $n^M$ being the unit 
normal to $\Sigma$.

\subsection{Bulk Perturbation}

Since the unperturbed background has $G_n$ symmetry, perturbations 
of the bulk geometry and those of the brane can be decomposed into 
scalar-type, vector-type and tensor-type components with respect to 
the transformation behavior under diffeomorphisms of $\K^n$, and 
components of different types do not couple in the Einstein 
equations\cite{Kodama.H&Sasaki1984}. Further, each component can be 
expanded in terms of the harmonic tensors of the same type on 
$\K^n$. In particular, for the scalar-type perturbation, the bulk 
metric perturbation 
$h_{MN}:=\delta g_{MN}$ is expressed in terms of the scalar 
harmonics $\SHB$ defined by
$
(\hat \triangle + k^2)\SHB=0
$,
as
\begin{equation}
h_{ab}=f_{ab}\SHB,\ 
h_{ai}=-rf_a \hat D_i\SHB,\ 
h_{ij}=2r^2(\bar\R\gamma_{ij}\SHB+ H_T\hat D_i\hat 
D_j\SHB),
\end{equation}
where $\hat D$ denotes the covariant derivative with respect to the metric 
$d\sigma_n^2$. We can construct two gauge-invariant combinations from the 
expansion coefficients $f_{ab}, f_a, \bar\R$ and $H_T$ as
\begin{equation}
F=\bar\R+\frac{1}{r}D^ar X_a,\quad
F_{ab}=f_{ab}+D_aX_b+D_bX_a,
\end{equation}
where $X_a=rf_a+r^2D_a H_T$.

>From the Einstein equations for the bulk we can show that these 
gauge-invariants are expressed as
\begin{equation}
r^{n-2}F=\frac{1}{2n}\left(\square
 -\frac{2}{\ell^2}\right)\Omega,\quad
r^{n-2}F_{ab}=D_aD_b\Omega
-\left(\frac{n-1}{n}\square-\frac{n-2}{n\ell^2}\right)\Omega g_{ab},
\label{MV:maxsymm}
\end{equation}
in terms of the master variable $\Omega$ satisfying the wave equation
\begin{equation}
\square \Omega -\frac{n}{r}Dr\cdot D\Omega
-\left(\frac{k^2-nK}{r^2}-\frac{n-2}{\ell^2}\right)\Omega
=0,
\end{equation}
where $\square=D^aD_a$.

\subsection{Brane Perturbation}

The intrinsic metric perturbation $\delta g_{\mu\nu}$ of the brane 
can be expanded in terms of the harmonics as
\begin{equation}
\delta g_{\tau\tau}=-2\alpha\SHB,\ 
\delta g_{\tau i}=a\beta \hat D_i\SHB, \ 
\delta g_{ij}=2a^2 (\R\SHB\gamma_{ij}+h_T \hat D_i\hat 
D_j\SHB_{ij}),
\end{equation}
and the perturbation of the brane position is determined by $ 
n_M\delta z^M=Z_\perp \SHB$, from which we can construct a 
gauge-invariant quantity $Y_\perp:=Z_\perp-X_\perp$. The standard 
gauge-invariants for the intrinsic metric perturbation are expressed 
in terms of the bulk gauge-invariants and $Y_\perp$ as
\begin{equation}
\Phi:=\R-{\dot a}\sigma_g=F+\frac{D_\perp 
r}{r}Y_\perp, \
\Psi:=\alpha-\Tdot{(a\sigma_g)}=
 -\frac{1}{2}F_{\para\para}-K^\tau_\tau Y_\perp,
\end{equation}
where $\sigma_g:={a}\dot h_T-\beta$ and 
$F_{\para\para}=F_{ab}u^au^b$ with $u^a$ being the unit tangent 
vector $\dot y^a$ of the brane.

The intrinsic matter perturbation of the brane is expanded as
\begin{eqnarray}
\delta T^\tau_\tau&=&-\delta\rho \SHB,\ 
\delta T^\tau_i=-a(\rho+p)(v-\beta)\hat D_i\SHB,\
\delta T^i_j \nonumber \\
&=& \delta p \SHB\delta^i_j
+ \pi_T \left(\frac{1}{k^2}\hat D^i\hat D_j\SHB
+\frac{1}{n}\delta^i_j\SHB\right).
\end{eqnarray}
We can construct the following three gauge-invariants from these 
perturbation variables other than the anisotropic stress 
perturbation $\pi_T$, which is gauge invariant by itself:
\begin{equation}
V:=k(v-{a}\dot h_T),\quad
\rho\Delta:=\delta \rho -{a}\dot\rho (v-\beta),\quad
\Gamma:=\delta p -c_s^2\delta\rho.
\end{equation}

\subsection{Junction Condition}

By expressing the perturbation of the junction condition 
(\ref{JunctionCondition}) in terms of the gauge-invariant quantities 
defined so far, we obtain the following four equations:
\begin{eqnarray}
&& 
\frac{{\bar\kappa}^2}{k^2-nK}a^n\rho\Delta
=-rD_\perp\Frac(\Omega/r)
-2a^{n-2}Y_\perp,
\label{JC:scalar:maxsymm1}\\
&&
{\bar\kappa}^2\frac{a^{n-1}}{k}(\rho+p)V
=\Tdot{(D_\perp\Omega)}+K^\tau_\tau\dot\Omega
+2a^{n-1}\Tdot{\left(a^{-1}Y_\perp\right)},
\label{JC:scalar:maxsymm2}\\
&&
\frac{1}{a}\Tdot{(aV)}=\frac{k}{a}\Psi
+\frac{k}{a}\frac{\Gamma+c_s^2\rho\Delta}{\rho+p}
-\frac{n-1}{n}\frac{k^2-nK}{ak}\frac{\pi_T}{\rho+p},
\label{JC:scalar:maxsymm3}\\
&& 
2\frac{k^2}{a^2}Y_\perp={\bar\kappa}^2\pi_T.
\label{JC:scalar:maxsymm4}
\end{eqnarray}
Among these, the first two give expressions for the intrinsic matter 
gauge-invariant variables $\Delta$ and $V$ in terms of the bulk 
variable $\Omega$ and $Y_\perp$. On the other hand, the rest give 
the boundary conditions on the latter, because the anisotropic 
stress perturbation $\pi_T$ and the entropy perturbation $\Gamma$ 
are not dynamical and are expressed in terms of other dynamical 
variables when the matter model is specified. In particular, when the 
anisotropic stress perturbation vanishes, they give the following boundary 
condition for the bulk master variable:
\begin{eqnarray}
&& \Tddot{\left[rD_\perp\Frac(\Omega/r)\right]}
+ (2 + nc_s^2)\frac{\dot 
a}{a}\Tdot{\left[rD_\perp\Frac(\Omega/r)\right]}
 +\left\{-n(1+w)(2n-2+nw)\Frac(D_\perp r/r)^2 \right. \nonumber \\
&& \quad \quad + \left.  c_s^2\frac{k^2-nK}{a^2}\right\}\left[rD_\perp 
\Frac(\Omega/r)\right]
{ -(n-1)(1+w)\frac{k^2}{a^2}\frac{D_\perp r}{r}\Omega} 
={\bar\kappa}^2a^{n-2}\Gamma.
\end{eqnarray}

\section{Behavior of Cosmological Perturbations}

In this section we examine the behavior of scalar perturbations of 
the brane using the formalism presented in the last section. We only 
consider the case $n=3$ and $\pi_T=0$.

\subsection{Low Energy Region}

If we put $\rho=\rho_0+\rho_m$ where $\kappa^2\rho_0/6=1/\ell^2$ and 
$\kappa^2=\bar\kappa^2/\ell$, in the low energy region for which 
$\rho_m/\rho_0\ll1$ and $k\ell/a\ll1$, the evolution equations of the 
brane universe are approximately given by 
\begin{equation}
H^2:=\Frac(\dot a/a)^2\simeq \frac{\kappa^2}{3}\rho_m - \frac{K}{a^2},\quad
\dot\rho_m=-3(1+w_m)\rho_m H,
\end{equation}
where $w_m=p_m/\rho_m$, and the junction condition is expressed as
\begin{eqnarray}
-(k^2-3K)a\Gamma &\simeq & \Tddot{(a^3\rho\Delta)}
+(2+3c_s^2)H\Tdot{(a^3\rho \Delta)} \nonumber \\
&+&\left[-\frac{3}{2}(1+w_m)\left(H^2+\frac{K}{a^2}\right)
+c_s^2\frac{k^2-3K}{a^2}\right](a^3\rho\Delta).
\end{eqnarray}

These equations are the same as those for the standard model. Hence 
the density perturbation of the brane universe 
in the low energy region behaves in the same way as that in the 
standard model. However, this does not immediately imply that the 
same result holds for other quantities such as the curvature 
perturbation, because we do not have the equations
\begin{equation}
\kappa^2\rho\Delta=2a^{-2}(k^2-3K)\Phi,\quad
\Phi+\Psi=0.
\end{equation}

\subsection{Superhorizon Perturbations}

For $K=0$ (a spatially flat model) the general solution for $\Omega$ 
is given by
\begin{equation}
\Omega(r,t)
={2l^2r}\int d\omega [A(\omega)J_0(m\ell/r)
+B(\omega)N_0(m\ell/r)]e^{-i\omega t},
\end{equation}
where $\omega^2=m^2+k^2$. By inserting this expression into the 
junction condition,  we obtain the following estimates for 
superhorizon perturbations with $k\ell/aH\ll1$ in the low energy 
region: 
\begin{eqnarray}
&& \Phi\simeq -\frac{2}{\pi}\int_{|\omega|<M}
d\omega B(\omega)e^{-i\omega t},\quad
Z\simeq\frac{5+3w_m}{3(1+w_m)}\Phi,\\
&& \Phi+\Psi \simeq \Order{h^2}\Phi,\quad
\frac{a^2}{2k^2}\kappa^2\rho\Delta
=\left(1+\Order{h^2}\right)\Phi, 
\end{eqnarray}
where $h=H\ell$, $M$ is a constant of the order $1/t$ and
$Z:=\Phi-aHV/k$ is the Bardeen parameter.
This together with the previous argument on $\Delta$ 
implies that for superhorizon modes in the low energy region 
the cosmological perturbation of the brane universe behaves 
in the same way as that of the standard model with no 
extra-dimension as far as the growing mode is concerned.
This confirms the same conclusion obtained by Koyama and 
Soda\cite{Koyama.K&Soda2000A} by a cruder estimate in the Gaussian 
normal coordinates.
Here note that $B(\omega)$ should have a $\delta$-function type peak 
at $\omega=0$ in order for $\Phi$ to approach a non-vanishing 
constant in the $t\rightarrow \infty$ limit. Further $B(\omega)$ 
should also have peaks for the values of $\omega$ corresponding to 
the time $t\sim 1/\omega$ at which $w_m$ changes.

In contrast, in the high energy region where $h\gg1$, we obtain the 
estimate $Z=\Order{h^2}\Phi$. Since $Z$ and $\Phi$ become constants 
of the same order for superhorizon perturbations in the standard 
models, this implies that the behavior of perturbations in the 
brane-world model is quite different from that in the standard model 
in the high energy region.



\end{document}